\documentclass[twocolumn,aps,pra,10pt,showpacs]{revtex4-2}
\usepackage{bm}
\usepackage{amsfonts}
\usepackage{amssymb}
\usepackage{amsmath} 

\begin{document}

\title{Universal uncertainty relations in classical and quantum theories of electromagnetism}
\author{Iwo Bialynicki-Birula}\email{birula@cft.edu.pl}
\affiliation{Center for Theoretical Physics, Polish Academy of Sciences\\
Aleja Lotnik\'ow 32/46, 02-668 Warsaw, Poland}
\author{Zofia Bialynicka-Birula}
\affiliation{Institute of Physics, Polish Academy of Sciences\\
Aleja Lotnik\'ow 32/46, 02-668 Warsaw, Poland}

\begin{abstract}
Sharp uncertainty relations restricting the values of variances in the position space and in the momentum (wavevector) space are derived. They have {\em the same form} $\Delta r\Delta k\ge 5/2$ in the classical theory of light beams, in the quantum theory of coherent light beams, and in the quantum theory of individual photons.
\end{abstract}
\maketitle

\section{Introduction}

In recent advanced experiments in nanophotonics \cite{wu,lukin,adv,tong,pra1,pra2} the limitations imposed by the uncertainty relations play a significant role in reaching extreme optical field confinement. In the present work we clarify the role of the uncertainty relations in all such experiments by deriving sharp uncertainty relations for light, no matter whether one employs classical or quantum description. Our uncertainty relations imply that better focusing must lead always to a larger spread in the bandwidth, regardless whether one uses classical or quantum theory. We prove the uncertainty relation in three distinct cases: in the classical Maxwell theory, in the quantum theory of coherent light beams and in the quantum theory of single photons. Quite unexpectedly, the uncertainty relations are universal. In all three cases they have not only exactly the same form:
\begin{eqnarray}\label{ur}
\Delta r\Delta k\ge 5/2,
\end{eqnarray}
where $\Delta r$ and $\Delta k$ are square roots of the variances $\Delta r^2$ and $\Delta k^2$, but also the functions that saturate the inequality (\ref{ur}) are the same: they are obtained from the same variational principle. We use the reciprocal space of wavevectors rather than momentum space in order to eliminate the Planck constant from our uncertainty relations. This convention emphasizes the classical--quantum universality of our results. We believe that these findings add also a new perspective to the ongoing debate on the quantumness of the Heisenberg uncertainty relations \cite{heis,yang,nclas}.

Uncertainty relation in classical optics have been discussed previously \cite{af,mm}. What distinguishes the approach adopted in the present work is the unified treatment of the classical and quantum descriptions, which leads to the same uncertainty relation (\ref{ur}) in both cases.

In our earlier works \cite{prl,hph}, we derived the uncertainty relations for photons based on different definitions of the dispersion in space. These derivations yielded the lower bounds in the uncertainty relations equal to 4 and $1+\sqrt{5}/2$. Instead, in the present work we use systematically the energy density to characterize the spatial extent of the field. The classical-quantum relationship in the uncertainty relations has been analyzed in detail in the review article \cite{dd}. Here we significantly extend this analysis by calculating explicitly not only the lower bound but also the functions which saturate the uncertainty relations.

The main theoretical tool used in our derivations is the Riemann-Silberstein (RS) vector ${\bm F}(\bm{r},t)$ \cite{ls,pwf0,pwf,rs},
\begin{eqnarray}\label{rs}
{\bm F}(\bm{r},t)=\frac{{\bm D}({\bm r},t)}{\sqrt{2\epsilon_0}}+i\frac{{\bm B}({\bm r},t)}{\sqrt{2\mu_0}}.
\end{eqnarray}
 
The description in terms of the RS vector is universal and applies to all three cases considered here, although its physical interpretation differs in each case.

\section{Uncertainty relation for classical electromagnetic waves}

Maxwell equations expressed in terms of the RS vector take the form,
\begin{eqnarray}\label{meq}
i\partial_t{\bm F}(\bm r,t)=c{\bm\nabla}\times{\bm F}(\bm r,t).
\end{eqnarray}
The general solution of this equation is given by the Fourier integral,
\begin{eqnarray}\label{rs1}
{\bm F}(\bm r,t)=\int\frac{d^3k}{(2\pi)^{3/2}}{\bm e}(\bm k)\nonumber\\
\left[f_+({\bm k})e^{i\bm k\cdot\bm r-i ckt}+f_-^*({\bm k})e^{-i\bm k\cdot\bm r+i ckt}\right],
\end{eqnarray}
where $k=\sqrt{k_x^2+k_y^2+k_z^2}$, the indices $\pm$ label waves of positive/negative helicity (left-handed and right-handed circular polarization) and
${\bm e}(\bm k)$ is the normalized polarization vector,
\begin{eqnarray}\label{ee}
{\bm e}(\bm k)=\frac{1}{\sqrt{2k^2(k_x^2+k_y^2)}}
\left[\begin{array}{c}
-k_xk_z+i k_y k\\
-k_yk_z-i k_x k\\
k_x^2+k_y^2
\end{array}\right]={\bm e}^*(-\bm k).
\end{eqnarray}
We will also need the Fourier representation of the RS vector,
\begin{eqnarray}\label{rs2}
{\bm\tilde F}(\bm k,t)=\left[{\bm e}(\bm k)f_+({\bm k})e^{-i ck t}\!+\!{\bm e}^*(\bm k)f_-^*(-{\bm k})e^{i ckt}\right].
\end{eqnarray}

The time-component $\rho({\bm r},t)$ of a conserved four-vector current could be used to characterize the distribution in space. This approach formed the basis of the uncertainty relations for relativistic electrons \cite{elur}. The electromagnetic field, however, does not possess a conserved four-current. A natural substitute for $\rho({\bm r},t)$ is the field energy density ${\bm F}^*(\bm r,t)\!\cdot\!{\bm F}(\bm r,t)$ whose variance $\Delta r^2$ we use as a measure the spatial spread of the field. Since we are interested in the smallest possible spatial extent, we set $t=0$. At other times the extent of the wave packet is larger because the mean-square radius grows quadratically with time,
\begin{eqnarray}\label{wp}
\frac{d^2}{dt^2}\frac{\int\!d^3r\,{\bm r}^2{\bm F}^*({\bm r},t)\!\cdot\!{\bm F}({\bm r},t)}{\int\!d^3r{\bm F}^*({\bm r},t)\!\cdot\!{\bm F}({\bm r},t)}=2 c^2.
\end{eqnarray}
The variance in position space is therefore defined by,
\begin{eqnarray}\label{vr}
\Delta r^2=\frac{\int\!d^3r\,{\bm r}^2{\bm F}^*(\bm r)\!\cdot\!{\bm F}(\bm r)}{\int\!d^3r{\bm F}^*(\bm r)\!\cdot\!{\bm F}(\bm r)},
\end{eqnarray}
where all field vectors are evaluated at $t=0$. By analogy, we define the variance in wavevector space as,
\begin{eqnarray}\label{vk}
\Delta k^2=\frac{\int\!d^3k\,{\bm k}^2{\bm\tilde F}^*(\bm k)\!\cdot\!{\bm\tilde F}(\bm k)}{\int\!d^3k{\bm\tilde  F}^*(\bm k
)\!\cdot\!{\bm\tilde  F}(\bm k)}.
\end{eqnarray}
By the Plancherel theorem, the denominators in (\ref{vr}) and (\ref{vk}) are equal,
\begin{eqnarray}\label{eq}
N&=\int\!d^3r{\bm F}^*(\bm r)\!\cdot\!{\bm F}(\bm r)=\int\!d^3k{\bm\tilde F}^*(\bm k)\!\cdot\!{\bm\tilde F}(\bm k)\nonumber\\
&=\int\!d^3k\left(|f_+({\bm k})|^2+|f_-({\bm k})|^2\right).
\end{eqnarray}
The equations determining the lowest bound in the uncertainty relation are obtained by varying the product $\Delta r^2\Delta k^2$ with respect to $f_\pm^*({\bm k})$. To derive these equations, we express $\Delta r^2$ in terms of the amplitudes $f_\pm({\bm k})$ by replacing the multiplication by $\bm r$ with differentiation of $f_\pm({\bm k})$ with respect to $\bm k$. We obtain,
\begin{widetext}
\begin{eqnarray}\label{ff1}
\Delta r^2=\frac{1}{N}\int\!d^3k\,f_+^*({\bm k})\left[\frac{1}{k_\perp^2}f_+({\bm k})+\frac{2i k_z}{kk_\perp^2}\left(k_x\partial_{k_y}f_+({\bm k})-k_y\partial_{k_x}f_+({\bm k})\right)-\Delta_kf_+({\bm k})\right]+(f_+({\bm k})\to f_-({\bm k})),
\end{eqnarray}
\end{widetext}
where $\Delta_k$ denotes the Laplacian in wavevector space. The corresponding expression for $\Delta k^2$ is,
\begin{eqnarray}\label{ff2}
\Delta k^2=\frac{1}{N}\int\!d^3k\,k^2\,f_+^*({\bm k})f_+({\bm k})+(f_+({\bm k})\to f_-({\bm k})).
\end{eqnarray}
Since the functions $f_+({\bm k})$ and $f_-({\bm k})$ enter the expressions for $\Delta r^2$ and $\Delta k^2$ in exactly the same way, we first consider $f_+({\bm k})$ and we suppress the subscript $+$. The variational equation then becomes,
\begin{widetext}
\begin{eqnarray}\label{ur1}
N\frac{\delta(\Delta r^2\Delta k^2)}{\delta f^*({\bm k})}=\Delta k^2\left[\!\frac{1}{k_\perp^2}-\frac{2k_z}{kk_\perp^2}
M_z-\Delta_k+\frac{\Delta r^2}{\Delta k^2} k^2-2\Delta r^2\right]f({\bm k})=0,
\end{eqnarray}
\end{widetext}
where $M_z=-i (k_x\partial_{k_y}-k_y\partial_{k_x})$ is the $z$-component on the angular momentum operator.
In spherical coordinates, the differential equation (\ref{ur1}) takes the form,
\begin{widetext}
\begin{eqnarray}\label{ur2}
\left[\!-\frac{1}{k^2}\partial_kk^2\partial_k
\!-\!\frac{1}{k^2\sin\theta}\partial_\theta\sin\theta\,\partial_\theta
\!+\!\frac{1\!-\!\partial_\phi^2\!+\!2i\cos\theta\partial_\phi}{k^2\sin^2\theta}\!+\!\frac{\Delta r^2k^2-2\gamma^2}{\Delta k^2}\!\right]\!\!f(k,\theta,\phi)\!=\!0,
\end{eqnarray}
\end{widetext}
where $\gamma^2=\Delta r^2\Delta k^2$. This equation is a special case of Eq.(23) in \cite{bb4} corresponding to the helicity parameter $h=1$. The general solution of this equation can be obtained by the separation of variables. Assuming the solution in the product form $f(k,\theta,\phi)=\Upsilon(k)\Theta(\theta,\phi)$, we separate the angular and radial variables and obtain two separate equations one for $\Theta(\theta,\phi)$ and the other for $\Upsilon(k)$,
\begin{widetext} 
\begin{eqnarray}
\left[-\frac{1}{\sin\theta}\partial_\theta\sin\theta\,\partial_\theta
+\frac{1-\partial_\phi^2+2i\cos\theta\partial_\phi}{\sin^2\theta}\right]\Theta(\theta,\phi)
&=\lambda\Theta(\theta,\phi),\label{sep1}\\
\frac{1}{2}\left[\Delta k^2\left(-\partial_k^2-\frac{2}{\kappa}\partial_\kappa+\frac{\lambda}{k^2}\right)+\Delta r^2 k^2\right]\Upsilon(k)&=\gamma^2 \Upsilon(k),\label{sep2}
\end{eqnarray}
\end{widetext}
where $\lambda$ is the separation constant. The equation for $\Theta(\theta,\phi)$ is the $s=1$ special case of the equation for spin-weighted spherical harmonics $_sY_{lm}$. Functions ot this type were first used in the theory of magnetic monopoles \cite{tamm} and later played an important role in general relativity \cite{gr,gr1}. Regular solutions of (\ref{sep1}) exist for $\lambda=j(j+1)$, where $j=1,2,\dots$. The equation for the radial part $\Upsilon(k)$ can be converted to a dimensionless form by the substitution $k=\kappa \Delta k/\sqrt{\Delta r\Delta k}$,
\begin{eqnarray}\label{sep3}
\frac{1}{2}\left[-\partial_\kappa^2-\frac{2}{\kappa}\partial_\kappa+\frac{j(j+1)}{\kappa^2}+\kappa^2\right]
\Upsilon(\kappa)=\gamma \Upsilon(\kappa),
\end{eqnarray}
where $\gamma=\Delta r\Delta k$. This is the radial eigenvalue equation for the three-dimensional harmonic oscillator. The lowest eigenvalue $\gamma=5/2$ (i.e. the ground state of the harmonic oscillator) is obtained for the smallest centrifugal potential, corresponding to $j=1$. This result was obtained in \cite{bb4}. Here, however, here we go further and give explicit formulas for the electric and magnetic fields which saturate the uncertainty relation. The radial ground state  function is,
\begin{eqnarray}\label{gs}
\Upsilon_g(\kappa)=\kappa\exp(-\kappa^2/2).
\end{eqnarray}                                                                                  
For $j=1$ there are three solutions $\Theta_\pm(\theta,\phi)$ and $\Theta_0(\theta,\phi)$ for the angular part,
\begin{eqnarray}\label{three}
\Theta_\pm(\theta,\phi)=(1\pm\cos\theta)e^{\pm i\phi},\quad \Theta_0(\theta,\phi)=\sin\theta.
\end{eqnarray}
The simplest RS vector is obtained for $\Theta_0(\theta,\phi)$. With this choice, the functions $f_\pm(\bm k)$ are,
\begin{eqnarray}\label{sub}
f_\pm({\bm k})=a\sqrt{k_x^2+k_y^2}\exp\left(-a^2 k^2/2\right),
\end{eqnarray}
where $a=\sqrt{\Delta r/\Delta k}$. In the next step we replace the components of the wave vector appearing in the polarization vector (\ref{ee}) by the corresponding spatial derivatives. This gives \cite{rs},
\begin{eqnarray}\label{rsf}
{\bm F}_\pm(\bm r,t)=\left[\begin{array}{c}\partial_x\partial_z+i\partial_y\partial_t/c\\
\partial_y\partial_z-i\partial_x\partial_t/c\\-\partial_x^2-\partial_y^2\end{array}
\right]{\mathcal F}_\pm(r,t),
\end{eqnarray}
where,
\begin{eqnarray}\label{caf}
{\mathcal F}_\pm(r,t)=\int\!\frac{d^3k}{4\pi\sqrt{\pi}k}\,
ae^{-a^2k^2/2}\left[e^{\pm i (\bm k\cdot\bm r-ckt)}\right].
\end{eqnarray}
This integral can be evaluated in closed form. Integrating first over the angular variables and then over $k$ we find,
\begin{eqnarray}\label{resf}
{\mathcal F}_\pm(r,t)
=\frac{1}{\sqrt{\pi}\,r}\int_0^\infty\!dk\,\sin(k r)\left[e^{-a^2k^2/2}e^{\mp i ckt}\right]\nonumber\\
=\frac{1}{\sqrt{2\pi}a r}\left[{\rm D}(l_+)+{\rm D}(l_-)\right]\pm\frac{i}{2\sqrt{2}ar}
\left[e^{-l_+^2}-e^{-l_-^2}\right]\!,
\end{eqnarray}
where $l_\pm=(r\pm ct)/(\sqrt{2}\,a)$ are the light-cone variables and ${\rm D}(l)=e^{-l^2}{\rm erfi}(l)$ is the Dawson function \cite{math}. The complete RS vector in position space is obtained by evaluating the derivatives of ${\mathcal F}_\pm(r,t)$. Although the resulting expressions are lengthy, they involve only Dawson functions and the exponential functions of the squared light-cone variables $l_\pm^2$. Superpositions of the two functions ${\mathcal F}_+(r,t)$ and ${\mathcal F}_-(r,t)$ also saturate the uncertainty relation because both functions $f_\pm({\bm k})$ satisfy the same equation (\ref{ur2}). The simplest such superposition is the difference ${\mathcal F}_+(r,t)-{\mathcal F}_-(r,t)$ because Dawson functions cancel leaving only exponentials. Discarding an irrelevant multiplicative constant we write,
\begin{eqnarray}\label{simple}
{\mathcal F}(r,t)=\frac{1}{r}\left(e^{-l_+^2}-e^{-l_-^2}\right). 
\end{eqnarray}
The corresponding RS vector has the following components:
\begin{widetext}
\begin{eqnarray}\label{simple1}
F_x=\frac{2 e^{-\frac{c^2 t^2+r^2}{2 a^2}}}{a^4 r^5}\Bigg(i \cosh \left(\frac{c r t}{a^2}\right)\,r \left(-i c t x z \left(3 a^2+2r^2\right)+c^2 t^2 r^2 y+r^4 y\right)\nonumber\\
\hspace{-2cm}+\sinh \left(\frac{c r t}{a^2}\right)
   \left(-3 a^4 x z-r^2 \left(a^2 (2 x z+i c t y)+c^2 t^2 x z+2 i c t r^2 y+r^2 x z\right)\right)\Bigg)\\
F_y=F_x \;{\rm with}\; \{x\to y,y\to -x\}, \;{\rm i.e.}\; F_y\;{\rm is\;obtained\;from}\;F_x\;{\rm by}\;90^\circ\;{\rm rotation},\\
F_z=\frac{e^{-\frac{c^2 t^2+r^2}{2 a^2}}}{a^4 r^5} \Bigg(\left(2 a^4 \left(r^2-3 z^2\right)-2
a^2 r^2 z^2+r^2 \left(r^2-z^2\right) \left(c^2 t^2+r^2\right)\right)\sinh
\left(\frac{c r t}{a^2}\right)\nonumber\\
-2 c r t \left(a^2 \left(r^2-3 z^2\right)+2
r^2 \left(r^2-z^2\right)\right) \cosh\left(\frac{c r t}{a^2}\right)\Bigg).
\end{eqnarray}
\end{widetext}
The RS vector built from the components $(F_x,F_y,F_z)$ obeys the complex form (\ref{meq}) of Maxwell equations.

At $t=0$ when the field has its minimal spread, it has only the magnetic component,
 \begin{eqnarray}\label{magr}
{\bm F}(\bm{r})=i \exp\left(-\frac{r^2}{2 a^2}\right)\left[\begin{array}{c}y\\-x\\0\end{array}
\right],
\end{eqnarray}
where an irrelevant constant multiplier has been omitted. The Fourier transform of $\bm{F}(\bm{r})$ (again up to an overall constant) is,
\begin{eqnarray}\label{magk}
\tilde{\bm{F}}(\bm{k})=i \exp{\left(-\frac{a^2k^2}{2}\right)}
\left[\begin{array}{c}k_y\\-k_x\\0
\end{array}\right]. 
\end{eqnarray}
Substitution of (\ref{magr}) and (\ref{magk}) into (\ref{vr}) and (\ref{vk}) gives,
\begin{eqnarray}\label{vrk}
\frac{\int\!d^3r\,{\bm r}^2{\bm F}^*(\bm r)\!\cdot\!{\bm F}(\bm r)}{\int\!d^3r{\bm F}^*(\bm r)\!\cdot\!{\bm F}(\bm r)}=\frac{5 a^2}{2},\\
\frac{\int\!d^3k\,{\bm k}^2{\bm\tilde F}^*(\bm k)\!\cdot\!{\bm\tilde F}(\bm k)}{\int\!d^3k{\bm\tilde  F}^*(\bm k )\!\cdot\!{\bm\tilde  F}(\bm k)}=\frac{5}{2a^2},
\end{eqnarray}
which confirms our uncertainty relation (\ref{ur}).
 
\section{Uncertainty relation for coherent states of the quantized electromagnetic field}

Quantization proceeds according to the standard rules. The c-number amplitudes $f_\pm({\bm k})$ and $f_\pm^*({\bm k})$ are replaced by annihilation and creation operators. These operators satisfy the canonical commutation relations,
\begin{eqnarray}\label{cr}
[a_+({\bm k}),a^\dagger_+({\bm k}')]=k\delta^{(3)}({\bm k}-{\bm k}')=[a_-({\bm k}),a^\dagger_-({\bm k}')].
\end{eqnarray}
Strictly speaking, a dimensional factor $\sqrt{\hbar c}$ should be included in these replacements in order to obtain the proper commutation relations for the field operators,
\begin{eqnarray}\label{can}
[\hat{B}_i({\bm r},t),\hat{D}_j({\bm r}',t)]=i\hbar\epsilon_{ijk}\partial_k\delta^{(3)}({\bm r}-{\bm r}').
\end{eqnarray} 
However, the factor $\sqrt{\hbar c}$ does affect our uncertainty relations because it cancels in the definitions of the variances (\ref{vr}) and (\ref{vk}).

As in the classical theory, the energy density operator $\hat{\mathcal{E}}({\bm r},t)$ plays the central role in the quantized theory,
\begin{eqnarray}\label{enmom}
\hat{\mathcal{E}}({\bm r},t)&=:\!\hat{\bm F}^\dagger(\bm{r},t)\cdot\hat{\bm F}^\dagger(\bm{r},t)\!:.
\end{eqnarray}
The double colons denote normal ordering of the creation and annihilation operators. We construct a general coherent state vector $|{\rm coh}\rangle$ by applying the unitary Glauber displacement operator \cite{rg} to the vacuum state vector,
\begin{eqnarray}\label{coh}
|{\rm coh}\rangle=e^{a^\dagger_++a^\dagger_--a_+-a_-}|0\rangle,
\end{eqnarray}
where $a^\dagger_\pm$ and $a_\pm$ are creation and annihilation operators constructed from arbitrary wave functions $f_\pm({\bm k})$,
\begin{eqnarray}\label{aad}
a^\dagger_\pm=\!\int\!\frac{d^3k}{k}f_\pm({\bm k})a^\dagger_\pm({\bm k}),\;\, a_\pm=\!\int\!\frac{d^3k}{k}f^*_\pm({\bm k})a_\pm({\bm k}).
\end{eqnarray}
The definitions of the second moments retain the form (\ref{vr}) and (\ref{vk}), but the classical expressions involving the RS vectors are now replaced by expectation values of the field operators in the coherent state,
\begin{eqnarray}
\Delta r^2=\frac{\int\!d^3r\langle{{\rm{coh}}|{\bm r}^2:{\hat{\bm F^\dagger}}({\bm r})\cdot{\hat{\bm F}}(\bm r):|{\rm{coh}}\rangle}}{{\int\!d^3r\langle{\rm{coh}}|:{\hat{\bm F^\dagger}}(\bm r)\cdot{\hat{\bm F}}(\bm r):|{\rm{coh}}\rangle}},\label{vrq}\\
\Delta k^2=\frac{\int\!d^3k\langle{{\rm{coh}}|{\bm k}^2:{\hat{\bm F^\dagger}}({\bm k})\cdot{\hat{\bm F}}(\bm k):|{\rm{coh}}\rangle}}{{\int\!d^3k\langle{\rm{coh}}|:{\hat{\bm F^\dagger}}(\bm k)\cdot{\hat{\bm F}}(\bm k):|{\rm{coh}}\rangle}}.\label{vkq}
\end{eqnarray}
Evaluating the expectation values of normally ordered products of field operators in a coherent state amounts to replacing the creation and annihilation operators by the corresponding functions $f_\pm({\bm k})$ and $f_\pm^*({\bm k})$. Consequently, the definitions (\ref{vrq}) and (\ref{vkq}) reduce to the classical definitions (\ref{vr}) and (\ref{vk}).

Since the evaluation of the expectation value of field operators in coherent states amounts to replacing all creation and annihilation operators by the corresponding wave functions $f_\pm({\bm k})$ and $f_\pm^*({\bm k})$, the definitions (\ref{vrq}) and (\ref{vkq}) are the same as the definitions (\ref{vr}) and (\ref{vk}) in the classical theory.

Therefore, the uncertainty relations for coherent states of the quantized electromagnetic field have the same form (\ref{ur}) as in the classical theory.

\section{Uncertainty relation for individual photons}

In our previous publication \cite{bb4} we derived the uncertainty relation for {\em all} massless particles,
\begin{eqnarray}\label{massless}
\Delta r\Delta k\ge 1+\sqrt{1/4+2 h},
\end{eqnarray}
where $h$ is the modulus of the helicity. For photons, $h=1$, this relation coincides with (\ref{ur}). The derivation of the formula (\ref{massless}) was based on the description of the wave functions in terms of relativistic spinors. In the present work, we have chosen the description in terms of the RS vector because it provides a unified treatment of the three cases considered here.

States of photons with positive and negative helicity are described by two wave functions $\bm F_+$ and $\bm F_-$, both of each can be obtained from the RS vector \cite{pwf0,pwf,rs}. Since photons, like all elementary particles, have positive energy, only the positive-frequency part of the RS vector (\ref{rs1}) represents a physical photon wave function. The wave function of photons with positive helicity is,
\begin{eqnarray}\label{wfp}
{\bm F_+}(\bm r,t)=\int\!\frac{d^3k}{(2\pi)^{3/2}}{\bm e}(\bm k)f_+({\bm k})e^{i\bm k\cdot\bm r-i ckt}.
\end{eqnarray}
The wave function of photons with negative helicity is obtained from the RS vector by complex conjugation of the negative frequency part,
\begin{eqnarray}\label{wfm}
{\bm F_-}(\bm r,t)=\int\!\frac{d^3k}{(2\pi)^{3/2}}{\bm e^*}(\bm k)f_-({\bm k})e^{i\bm k\cdot\bm r-i ckt}.
\end{eqnarray}
As shown by Wigner \cite{wig}, the two wave functions ${\bm F_-}(\bm r,t)$ are two different representations and are different geometrical objects. Under rotations, they transform in the same way,
\begin{eqnarray}\label{rot}
{\bm F}'={\bm F}\cos\varphi+{\bm n}\times{\bm F}\sin\varphi
+{\bm n}\left({\bm n}\cdot{\bm F}\right)(1-\cos\varphi),
\end{eqnarray}
where ${\bm n}$ is the rotation axis and $\varphi$ is the rotation angle.
Their transformation laws under Lorenz boosts, however, differ in the sign of the second term in \cite{pwf} their transformation properties under Lorentz transformations ,
\begin{eqnarray}\label{lorentz}
{\bm F}'\!=\!{\bm F}\cosh\psi\!\mp\! i{\bm n}\!\times\!{\bm F}\sinh\psi
\!+\!{\bm n}\left({\bm n}\cdot{\bm F}\right)(1\!-\!\cosh\psi)\!,
\end{eqnarray}
where ${\bm n}={\bm v}/|{\bm v}|$ and $\psi={\rm{arctanh}}(|{\bm v}|/c)$.

The helicity of photons differs in an important way from the helicity of massive particles. For massive particles, helicity states mix under rotations and Lorentz transformations, whereas for photons they transform separately. In both cases, however, helicity contributes $\pm\hbar$ to the total angular momentum, as discussed in detail in \cite{sep,aiello}.

For photons, the uncertainty relation (\ref{ur}) follows directly from our results for the classical electromagnetic field. The calculations are identical to those leading to the formula (\ref{resf}). In the present case, however, the amplitudes $f_\pm({\bm k})$ are photon wave functions in Fourier space rather than amplitudes of the classical electromagnetic wave. We, therefore cannot use the simple solution (\ref{simple}) because it contains both: positive and negative energy contributions. Instead, we must consider separately the positive helicity ${\bm \Psi}_+(\bm r,t)$ and negative helicity ${\bm \Psi}_-(\bm r,t)$  photon wave functions,
\begin{eqnarray}\label{rsfpm}
{\bm\Psi}_\pm(\bm r,t)=\left[\begin{array}{c}\partial_x\partial_z\pm i\partial_y\partial_t/c\\
\partial_y\partial_z\mp i\partial_x\partial_t/c\\-\partial_x^2-\partial_y^2\end{array}
\right]{\mathcal F}_\pm(r,t).
\end{eqnarray}
The expressions obtained from this formula, unlike those for the electromagnetic field (\ref{simple1}), take several pages and will not be given here. To obtain the uncertainty relations for photons it is sufficient to consider the photon wave function at $t=0$ when the extension of the wave packet is minimal. In this case all expressions are greatly simplified and the components of the photon wave function ${\bm\Psi}(\bm r,t)$ at $t=0$ are,
\begin{widetext}
\begin{eqnarray}\label{pwf}
&\Psi_x|_{t=0}=\mp\frac{y e^{-\frac{r^2}{2 a^2}}}{\sqrt{2} a^5}-\frac{x z \left(3
   a^2+r^2\right)}{\sqrt{\pi } a^4 r^4}+\frac{\sqrt{2} x z \left(3
   a^4+2 a^2 r^2+r^4\right) {\rm D}\left(\frac{r}{\sqrt{2} a}\right)}{\sqrt{\pi }a^5 r^5},\\
&\Psi_y|_{t=0}=\pm \frac{x e^{-\frac{r^2}{2 a^2}}}{\sqrt{2} a^5}-\frac{y z \left(3
   a^2+r^2\right)}{\sqrt{\pi } a^4 r^4}+\frac{\sqrt{2} y z \left(3
   a^4+2 a^2 r^2+r^4\right) {\rm D}\left(\frac{r}{\sqrt{2} a}\right)}{\sqrt{\pi }a^5 r^5},\\
&\Psi_z|_{t=0}=\frac{a r \left(a^2+r^2\right) \left(r^2-z^2\right)-2 a^3 r z^2}{\sqrt{\pi } a^5
   r^5}\frac{-\sqrt{2}
\left(a^4 r^2-3 a^4 z^2-2 a^2 r^2 z^2+r^6-r^4 z^2\right){\rm D}\left(\frac{r}{\sqrt{2} a}\right)}{\sqrt{\pi } a^5 r^5}.  
\end{eqnarray}
\end{widetext}
The substitution of these functions into (\ref{vr}) again results in $\Delta r^2=5 a^2/2$. Similarly, substituting the Fourier transform ${\tilde{\bm F}}(\bm k,0)$ into (\ref{vk}) gives $\Delta k^2=5 /(2 a^2)$. This confirms our uncertainty relation (\ref{ur}) for the photon wave functions. As expected, the lower bound in the uncertainty relation for photons is larger than that for massive particles, because photons are more difficult to localize in phase space. 
\section{Conclusions}

The main conclusion of this work is that the uncertainty relations in the theories of electromagnetism characterize wave properties, rather than quantum properties. They can be derived without {\em any reference} to operators acting in the Hilbert space of quantum states. The fact that the uncertainty relations have the same form for classical electromagnetic waves and for photon wave functions is the best proof of this universality. To emphasize this property, we have eliminated completely the Planck constant from uncertainty relations by using the wave vector instead of momentum vector. By invoking the relation between the photon wave vector and the photon momentum $\bm p=\hbar\bm k$, our uncertainty relation can be written in the quantum-mechanical form $\Delta r\Delta p\ge 5/2\hbar$. The probabilistic nature of wave functions plays the significant role in the {\em physical interpretation} of uncertainty relations in quantum theory, however the mathematical derivation of these relations does not require the formal apparatus of quantum theory, such as state vectors in Hilbert space, operators or expectation values.

\end{document}